\documentstyle[twocolumn,prb,aps]{revtex}
\renewcommand{\narrowtext}{\begin{multicols}{2} \global\columnwidth20.5pc}

\begin{document}
\newcommand{\be}{\begin{equation}}
\newcommand{\ee}{\end{equation}}
\newcommand{\nt}{\narrowtext}

\noindent
Comment on $\bf Algebraic$  $\bf Fermi$ $\bf Liquid$ $\bf from$ 
$\bf Phase$ $\bf Fluctuations \dots$
% $\bf Topological$ $\bf Fermions$, $\bf Vortex$ $\bf Berryons$, $\bf and$ $QED$ $\bf theory$ $\bf of$ $\bf Cuprate$ $\bf Superconductors$

In a recent Letter [1], M. Franz and Z. Tesanovic predicted a Luttinger-like behavior in the pseudogap phase
of underdoped cuprates. This conclusion was drawn on the basis
of the following propositions: 1) the low-energy properties of the 
pseudogap phase are described by the $QED_3$-like effective theory
formulated in terms of massless "topological" fermions $\psi$ and a gauge field $A_\mu$
representing the effect of fluctuating vortices ("berryons");
2) the true physical electron propagator is given by the gauge-invariant amplitude
\be
G_0(x)=<\psi(x)\exp(-i\int_\Gamma A_\mu(z)dz^\mu){\overline \psi}(0)> 
\ee
with the contour $\Gamma$ chosen as the straight line between the end points; 3) 
the amplitude (1) demonstrates a power-law decay (here $\Lambda$ is an 
upper energy-momentum cutoff)
\be
G_0(x)\propto {{\hat x}/\Lambda^\eta|x|^{3+\eta}}
\ee
controlled by a positive anomalous dimension $\eta$.

Considering their largely heuristic nature,
the status of the propositions 1) and 2) is more difficult to ascertain.
However, the validity of 3) can be checked directly, and it appears to be merely incorrect.

The value of the anomalous dimension $\eta_1=16/3\pi^2N$ quoted in Ref.[1] without derivation
was later obtained in Ref.[2] where the authors substituted Eq.(1) with the so-called Brown's function
\be
G_1(x)={<\psi(x){\overline \psi}(0)>\over <\exp(i\int_\Gamma A_\mu(z)dz^\mu)>} 
\ee
which was claimed to be truly gauge invariant and, therefore,
identical to Eq.(1), since the two functions coincide in the axial gauge
($A_\mu(z)x_\mu=0$).

In contrast, a direct calculation of Eq.(1) yields a negative 
value $\eta_0=-32/3\pi^2N$ not only in the above axial gauge  
but also in the standard covariant [3] and the 
(considered even more reliable) radial ($A_\mu(z) z_\mu=0)$ [4] ones.
This result was obtained with the use of both the simple 
cut-off and the dimensional regularization schemes.

As for Eq.(3), the latter appears to be no different from any other function
\be
G_\xi(x)={<\psi(x)\exp(i(\xi-1)\int_\Gamma A_\mu(z)dz^\mu){\overline \psi}(0)>\over
<\exp(i\xi\int_\Gamma A_\mu(z)dz^\mu)>} 
\ee
which all 1) coincide with Eq.(1) in the axial gauge, 2) are
independent of the gauge parameter within the class of covariant gauges
(which fact was incorrectly interpreted in Ref.[2] as a sign of their absolute gauge-invariance), 
but, being given by a ratio of the expectation values of two gauge-variant operators, 
3) are not truly gauge-invariant, except for $\xi=0$ which corresponds to Eq.(1).
Indeed, in both the axial and the radial gauges
the anomalous dimension of Eq.(4) equals $\eta_0$ for any $\xi$, whereas
in the covariant gauge it shows an explicit $\xi$-dependence,
$\eta_\xi=16(3\xi-2)/3\pi^2N$ [3], thus demonstrating the lack of gauge invariance for any $\xi\neq 0$ 
including the case of $G_1(x)$ given by Eq.(3).

Taken at its face value, the negative anomalous 
dimension of the function $G_0(x)$ (which is the only gauge-invariant one amongst
the entire set (4)) disqualifies it (let along any gauge-variant 
function $G_\xi(x)$ with $\xi\neq 0$) from being
a sound candidate to the role of the physical electron propagator, 
since in the effective $QED_3$-like models 
the repulsive electron interactions are expected
to result in suppression of such an amplitude.

Furthermore, the same negative value $\eta_0$ was found for Eq.(1)
with the contour $\Gamma$ chosen as a pair of semi-infinite (anti)parallel strings
between the end points and infinity [4], thereby suggesting that the sign of
$\eta_0$ may not be readily changeable by modifying the contour $\Gamma$ 
in Eq.(1) or replacing it with a properly (as opposed to arbitrarily) weighted sum over different contours. 

Thus, it appears that, so far, in the context 
of the massless $QED_3$ no truly gauge-invariant and firmly physically justified 
(as opposed to those concocted solely for the sake of the argument, regardless of their physical 
content) one-fermion amplitude demonstrating the power-law behavior (2) 
with a positive $\eta$ has been constructed.
Instead, a gauge-invariant alternative to Eq.(1) proposed in Ref.[3]
along the lines of the previous work on massive $QED_4$ (see Ref.[5] and references therein)
features a log-normal behavior $G_{phys}(x)\propto\exp(-\alpha\ln^2(\Lambda|x|)){\hat x}/|x|^3$
with $\alpha\sim 1/N$, thus indicating the possibility of a stronger-than-Luttinger suppression.

To conclude, the massless $QED_3$-like theory of the pseudogap phase 
has not yet provided a solid basis for the 
asserted Luttinger-like behavior of the electron spectral function
(much less a "natural explanation" of ARPES data claimed in Ref.[1]), 
regardless of its (still largely unsettled) experimental status.

This work was supported by NSF under Grant DMR-0071362.

$\bf D. V. Khveshchenko$\\
Physics and Astronomy, UNC-Chapel Hill, NC 27599

\newpage

\end{document}